\documentclass[aps,pre,reprint,superscriptaddress]{revtex4-2}
\usepackage{graphicx}
\usepackage{physics}
\usepackage{float}
\usepackage{amsmath}
\usepackage{amsfonts}
\usepackage{amssymb}

\begin{document}


\title{Automatic Learning of Topological Phase Boundaries}


\author{Alexander Kerr}
\email[E-mail corresponding author at: ]{ajkerr0@gmail.com}
\author{Geo Jose}
\author{Colin Riggert}
\author{Kieran Mullen}
\affiliation{Homer L. Dodge Department of Physics and Astronomy, The University of Oklahoma, \\ 440 W. Brooks St., Norman, OK 73019, USA}
\affiliation{Center for Quantum Research and Technology, The University of Oklahoma, \\ 440 W. Brooks Street, Norman, Oklahoma 73019, USA}


\date{\today}

\begin{abstract}
	
Topological phase transitions, which do not adhere to Landau's phenomenological model (i.e. a spontaneous symmetry breaking process and vanishing local order parameters) have been actively researched in condensed matter physics.  Machine learning of topological phase transitions has generally proved difficult due to the global nature of the topological indices.  Only recently has the method of diffusion maps been shown to be effective at identifying changes in topological order.  However, previous diffusion map results required adjustments of two hyperparameters: a data length-scale and the number of phase boundaries.  In this article we introduce a heuristic that requires no such tuning.  This heuristic allows computer programs to locate appropriate hyperparameters without user input.  We demonstrate this method's efficacy by drawing remarkably accurate phase diagrams in three physical models: the Haldane model of graphene, a generalization of the Su-Schreiffer-Haeger (SSH) model, and a model for a quantum ring with tunnel junctions.  These diagrams are drawn, without human intervention, from a supplied range of model parameters.

\end{abstract}


\maketitle


\section{Introduction}\label{sec:intro}

Topological quantum phase transitions (TQPTs) are transitions between quantum states of matter with different topological properties, typically indicated by a discontinuous change in a topological invariant like the Chern number.  TQPTs differ from classical and quantum phase transitions in that the former are described by Landau theory, based on the emergence of a local order parameter that is usually associated with a broken symmetry at the phase transition \cite{gj1, gj2, gj3}.  This order parameter is non-zero in the ordered phase and zero in the disordered phase. These ideas generalize to the case of quantum phase transitions, wherein the Landau functional is replaced by the action of the underlying quantum field theory. TQPTs do not conform to this paradigm: the distinction between different phases is not a broken symmetry, but rather the value of a topological index. These indices are fundamentally associated with global features of the quantum state and are not naturally described by local order parameters.  The general program of study for topological phase transitions in a given model with a finite number of tunable parameters involves computing the desired topological index as a function of the parameters and identifying regions in the parameter space that pertain to different indices. This requires prior knowledge of the relevant topological invariant.  Here we present an unsupervised machine learning algorithm that efficiently maps the boundary of a TQPT without any foreknowledge of the underlying topological invariant.

Machine learning programs have been previously established in physics using neural networks as a variational ansatz for many-body systems \cite{Carleo602, doi:10.7566/JPSJ.86.093001, doi:10.7566/JPSJ.87.074002}, modeling potential energy surfaces with near \textit{ab-initio} accuracy \cite{botu2017, Mueller_Hernandez_Wang_2020, Chan_Narayanan_2019, Behler_2016}, improved Monte-Carlo sampling \cite{PhysRevB.95.041101,PhysRevB.95.035105}, and beyond \cite{RevModPhys.91.045002, 10.1088/1361-648X/abb895}.  Within the field of condensed matter physics, the identification of phases of matter with machine learning has become an active area of research \cite{carrasquilla2017, Shiina2020, wang2016, wetzel2017}.  Classifying phases of matter (a supervised learning process) has been demonstrated with neural networks \cite{carrasquilla2017, Shiina2020}.  Perhaps more interestingly, other algorithms are known to learn order parameters in an unsupervised manner \cite{wang2016, wetzel2017}, meaning the identity of each phase was not explicitly labeled for the algorithm \textit{a priori}.  For example, computer programs were able to identify the magnetization of a spin lattice as the key feature defining the phase transition by simply analyzing sampled spin configurations without incorporating any model Hamiltonian or parameters into the code.

Recently, unsupervised methods have distinguished different topological phases of matter via manifold learning \cite{rodriguez2019, che2020}.  The principle behind using manifold learning is that different states sharing a global property may have a large Euclidean separation ($\mathbb{L}^2$ norm of the difference between their local variables) while lying on the same manifold in phase space.  Manifold learning tools are designed to correlate data points living on the same manifold by a non-linear manipulation of the data.  In this article we demonstrate the ability of a particular manifold learning routine, diffusion maps, to learn the boundaries of topological phase transitions.  Previous implementations of diffusion maps \cite{rodriguez2019, che2020} could only accurately draw phase diagrams after the number of boundaries were specified.  We present a heuristic in which the number of phase boundaries in a cross-section of parameter space is automatically determined by a computer program.  From this, phase diagrams are drawn without human input as a set of Hamiltonian parameters are swept.

We consider three cases.  The first is an experiment which illustrates how diffusion maps can identify classical XY spins of different winding numbers.  The second is applying the new heuristic to draw two well-known phase diagrams: an extension of the Su-Schreiffer-Haeger (SSH) model \cite{gj5} which exhibits a phase transition between phases labeled by different winding numbers as the hopping parameters are varied, and the Haldane model of graphene \cite{gj6} which involves a transition in the Chern number as a function of the hopping parameters and magnetic phases. The third example is to investigate a novel topological phase transition in the ground state of a single electron on a set of ideal, tunneling coupled rings.  As the tunneling between the rings is increased, the winding number of the single particle ground state can change abruptly.

\section{Automatic Phase Diagram Method}

\subsection{Diffusion Maps}

\begin{figure*}[t]
	\centering
	\includegraphics[width=6.9in]{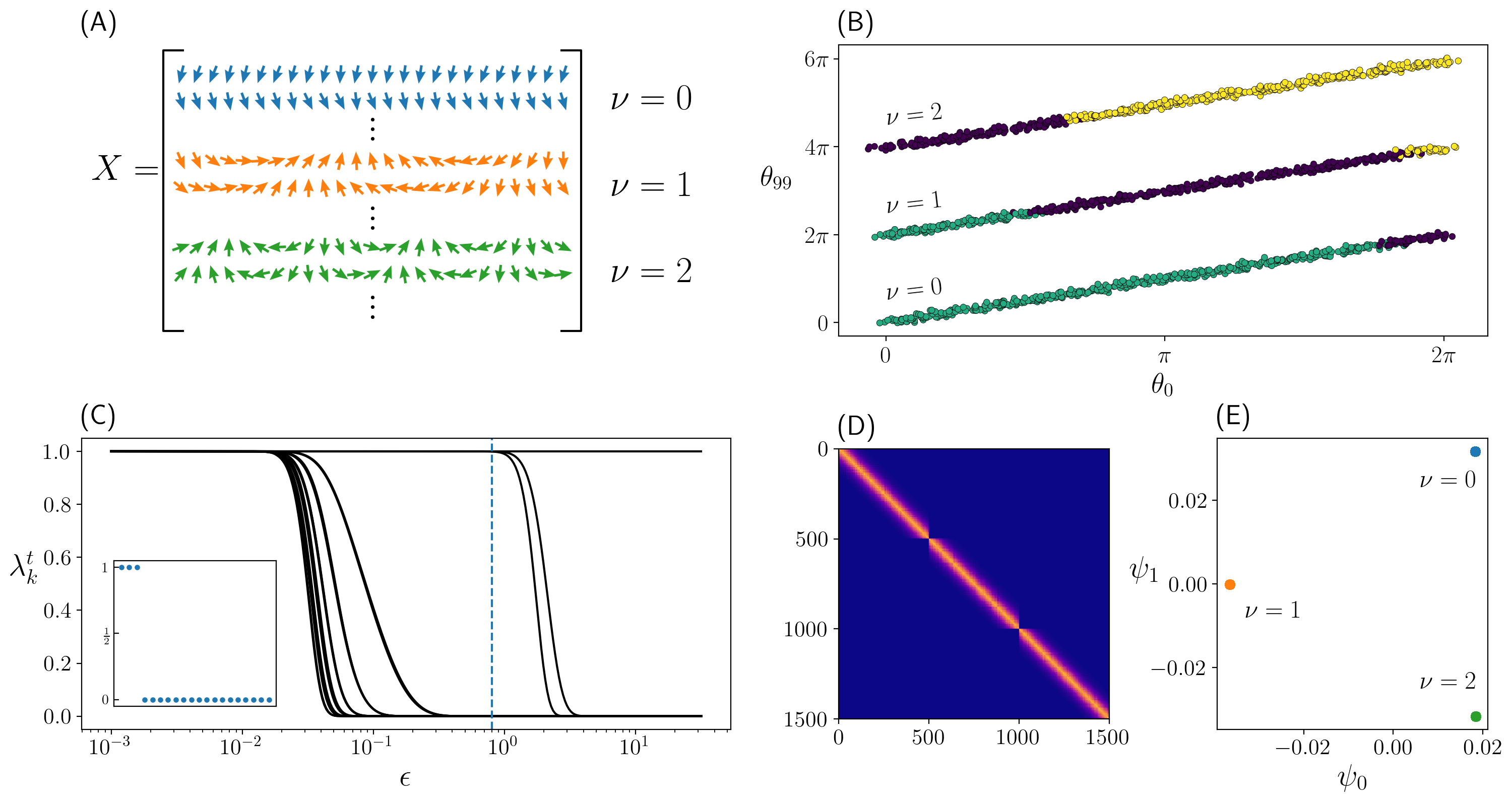}
	\caption{(A) A dataset is composed of classical XY spins with different winding numbers.  Points on a single manifold in parameter space can have a large pairwise separation, but a small diffusion separation as defined in Equation \ref{eq:diffdist}.  (B) The XY spins projected to two-dimensions.  The data is plotted as a function of the angles in the first and last lattice sites.  The data is stratified because of the presence of spin winding.  However, k-means clustering (color code) does not recognize the topological order of the full-dimensional dataset. (C) The 20 largest eigenvalues of the transition matrix $ P_{ij} $ for the data in Figure (A), as a function of the resolution parameter $ \epsilon $ in Equation \ref{eq:simmat} (note the semi-log scale).  The time scale here was chosen to be $ t=500 $.  $ P_{ij} $ is a probability matrix, and thus its eigenvalues ($ \lambda_k $) vary between 0 and 1.  The number of eigenvalues of order unity roughly correspond to the number of manifolds in the data.  Three eigenvalues survive for a large range of $ \epsilon $, indicating there are truly three sets of topologically distinct data.  The inset shows the eigenvalues for $ \epsilon = 0.8 $.  (D)  The similarity matrix of the data for $ \epsilon = 0.8 $, derived from Equation \ref{eq:simmat} using the $\mathbb{L}^{\infty}$ norm.  Much can be learned from visual inspection: 1) there are 3 manifolds in the data 2) the manifolds are well separated with weak inter-connections and strong intra-connections 3) each manifold is the same size 4) the manifolds have the same density.  (E)  The dataset transformed to two dimensions by diffusion maps.  The distribution of the data along the zeroth and first diffusion modes are shown.  Clearly the spin data in diffusion space is clustered according to their winding numbers, and k-means analysis automatically relates states of the same topological index.  The inter-cluster distance is much larger than the intra-cluster distance.  \label{fig:diffmap}}
\end{figure*}

Conventional clustering methods such as k-means \cite{kmeans} are not enough to discriminate the global indices of data.  The objective of k-means in particular is to minimize the squared Euclidean distance between data points and their cluster centers which inherently depends on local information.  In general, entries in a dataset may be well separated in parameter space despite sharing global properties.  An example is given in Figure \ref{fig:diffmap}(A): a series of classical XY spins on a 1D lattice.  A dataset is constructed containing spin-arrays with winding numbers $ \nu \in \{0,1,2\}$.  The dot product between two spin-arrays with the same winding number can be small (or very negative) even though they live on the same manifold in parameter space corresponding to their global property.  Therefore, the distance between two points is not enough information to determine if they contain the same topological index.  A naive k-means analysis (Figure \ref{fig:diffmap}(B)) fails to cluster the pure spin-data based on their winding number.  Meanwhile linear transformations, e.g. principal components analysis \cite{kmeans, ivezic}, cannot reduce this data down to its topological invariants.  The utility of manifold methods, like diffusion maps, is to cluster data in a non-linear manner.

Although data points on the same manifold (i.e. a given winding number) can be well separated, adjacent points typically lie on the same manifold.  Chains of locally-similar data can form globally-similar structures across large regions in the multi-dimensional phase space.  If allowed to jump across nearest-neighboring points, one could quickly diffuse through a single manifold.  From the speed of this diffusion, a new distance metric can be formulated which will separate data in a Euclidean manner.

Imagine a set of random walkers allowed to diffuse through a dataset.  The transition probability of a walker from data point $ \boldsymbol{x}_i $ to point $ \boldsymbol{x}_j $ ($ P_{ij} $) is the row-normalized Gaussian similarity measure
\begin{equation} \label{eq:probmat}
P_{ij} = \frac{\mathcal{K}_{ij}}{ \sum_j \mathcal{K}_{ij}}
\end{equation}
\begin{equation}\label{eq:simmat}
\mathcal{K}_{ij} = \exp\left(-\frac{|\boldsymbol{x}_i - \boldsymbol{x}_j |^2_p}{\epsilon}\right)
\end{equation}
where $ \boldsymbol{x}_i $ is the $i\text{th}$ data point in the given parameterization, $ \epsilon $ is a resolution hyperparameter setting a length scale, and $ |\cdot|_p $ denotes the $\mathbb{L}^p$ norm.  Note that while a data point $ \boldsymbol{x}_i $ is multi-dimensional in general, the matrix $ P_{ij} $ has dimensions $ N \times N $ where $ N $ is the number of points in the dataset.  These walkers are likely to hop between nearest-neighbors while jumps across large distances are exponentially suppressed.  We introduce the eigenvalues $ \lambda_k $ and eigenvectors $\psi_k$ of $ P_{ij} $ defined by
\begin{equation}\label{eq:eigendef}
P_{ij} \psi_k = \lambda_k \psi_k
\end{equation}
The diffusion distance $ D(\boldsymbol{x}_i, \boldsymbol{x}_j) $ between any two points $\boldsymbol{x}_i$ and $\boldsymbol{x}_j$ in the data set after $ t $ time steps is defined by \cite{COIFMAN20065}
\begin{equation}\label{eq:diffdist}
D^2_t(\boldsymbol{x}_i, \boldsymbol{x}_j) = \sum_k \lambda_k^{2t} \left[\psi_{k}(\boldsymbol{x}_i) -  \psi_{k}(\boldsymbol{x}_j)\right]^2
\end{equation}
where $ \psi_{k}(\boldsymbol{x}_i) $ is the $ i $th component of the $ k $th eigenvector.  Data points separated by a small diffusion distance $ D_t $ are understood to belong to the same manifold because $ D_t $ is small when there are many paths comprised of short-ranged hops connecting them.  Thus, the diffusion distance is a useful metric for the global-similarity of data points and is used to delineate manifolds.  Diffusion distance is not typically calculated explicitly however.  Instead, the original data set $ \{\boldsymbol{x}_{i} \}$ (Figure \ref{fig:diffmap}(A)) is transformed:
\begin{equation}\label{eq:transformdata}
\boldsymbol{y}_i = \left\{ \lambda_0^t \psi_0 (\boldsymbol{x}_i), \lambda_1^t \psi_1 (\boldsymbol{x}_i), \, ... \, , \lambda_{N-1}^t \psi_{N-1} (\boldsymbol{x}_i) \right\}
\end{equation}
where the eigenvalues are in descending order.  With this transformed data $ \{\boldsymbol{y}_{i} \}$, more routine data analysis can be applied.  This is because the Euclidean distance between the transformed points corresponds to the diffusion distance between the original entries \cite{COIFMAN20065}.  Another nice feature of this process is that the distance metric in Equation \ref{eq:simmat} can be generalized.  A physical picture for why diffusion maps work is to interpret $ P_{ij} $ as the dynamical matrix of a set of point masses connected by springs with a stiffness proportional to $ K_{ij} $.  A data manifold corresponds to an eigenmode that moves in phase due to their strong interconnections.

Because the transformation in Equation \ref{eq:transformdata} preserves the manifold information of the original dataset, it is understood that only a few of the largest $ \lambda_k $ contribute to the calculation of $ D_t(\boldsymbol{x}_i, \boldsymbol{x}_j) $.  The exact number of surviving eigenvalues is determined by the connectivity of the data which in turn depends on the resolution hyperparameter in Equation \ref{eq:simmat}.  Figure \ref{fig:diffmap}(C) shows the largest eigenvalues corresponding to the XY spin data as a function of the resolution.  When $ \epsilon $ is small, there are many surviving eigenvalues attributed to the large effective separation between the data points.  When this scale is small enough, individual data points become isolated and register as their own manifold.  As $ \epsilon $ is increased to the inherent length scale of the data, the true structure is elicited by the number of long-surviving eigenvalues.  When $ \epsilon $ is tuned very high, all but one eigenvalue vanish due to the entire dataset appearing as a single tightly-bound cluster.  Indeed, just three eigenvalues remain across a large range of resolution values hinting that the data is truly structured in three separate manifolds.  The similarity matrix for a select resolution is shown in Figure \ref{fig:diffmap}(D), containing even more information about the connectivity of the data.  The $\mathbb{L}^{\infty}$ norm was used to derive this similarity (Equation \ref{eq:simmat}) due to its ability to encode topological information \cite{che2020}.  The manifold form of the data can be visualized in diffusion space (Figure \ref{fig:diffmap}(E)).  Since the diffusion distance of the original data is roughly equated to the Euclidean distance between the transformed data points, the transformed data appears in three tightly-bound clusters.  Finally, k-means clustering can be applied to the diffusion modes to identify topological structure.  If the data in a set is ordered, then changes in the k-means cluster labels are interpreted as a phase boundary.

\subsection{Optimization of Resolution Hyperparameter}\label{sec:eps-choice}

Before diffusion maps can find phase boundaries in physical models, the resolution $ \epsilon $ of the data in Equation \ref{eq:simmat} must be chosen.  Conventionally this is performed by inspecting the probability eigenspectrum as a function of $ \epsilon $ (as in Figure \ref{fig:diffmap}(B)) and selecting a value for $ \epsilon $ that captures the manifold structure of the data.  The degeneracy of eigenvalues close to unity corresponds to the number of manifolds.  However, this process is ill-defined as it is generally unclear which resolution scale ``best" illustrates the global arrangement of the data.  In addition, this process must precede every experiment.

We introduce a heuristic that allows computer programs to automatically choose a resolution hyperparameter, without user input.  We seek the resolution that minimizes an adjusted mean squared distance between the similarity matrix in Equation \ref{eq:simmat} ($ \mathcal{K}(\epsilon) $) and the ``ideal'' similarity matrix ($ \mathcal{K}^{\text{ideal}} $):

\begin{equation}\label{eq:mse}
\text{MSE}(n, \epsilon) = \frac{n-1}{n}\sum_{\ell=1}^{n} \frac{1}{|S_{\ell}|} \sum_{i \in S_{\ell}} \sum_j \left( \mathcal{K}_{ij}(\epsilon) - \mathcal{K}_{ij}^{\text{ideal}} \right)^2
\end{equation}


\begin{equation}
\mathcal{K}_{ij}^{\text{ideal}} = 
\begin{cases}
1, & i,j \text{ are contained in identical} \\
\; & \qquad \;  \text{k-means clusters} \\
0, & \text{otherwise}
\end{cases}
\end{equation}
where $n$ is the number of k-means clusters and $S_{\ell}$ is the set of points in cluster $\ell$.  We are defining the ideal similarity matrix as having a zero similarity between points of different clusters, and an intra-cluster similarity of one.  The cluster assignments for the ideal matrix are determined by a k-means clustering application \cite{kmeans} on the data in diffusion space.  With the form given in Equation \ref{eq:mse}, every cluster  carries the same weight in the squared difference between the target and functional similarity matrices, no matter the relative sizes of the clusters.  The factor of $ (n-1)/n $ cancels out some of the natural advantage of assigning a large number of clusters to the MSE.

\begin{figure}[t!]
	\centering
	\includegraphics[width=3.3in]{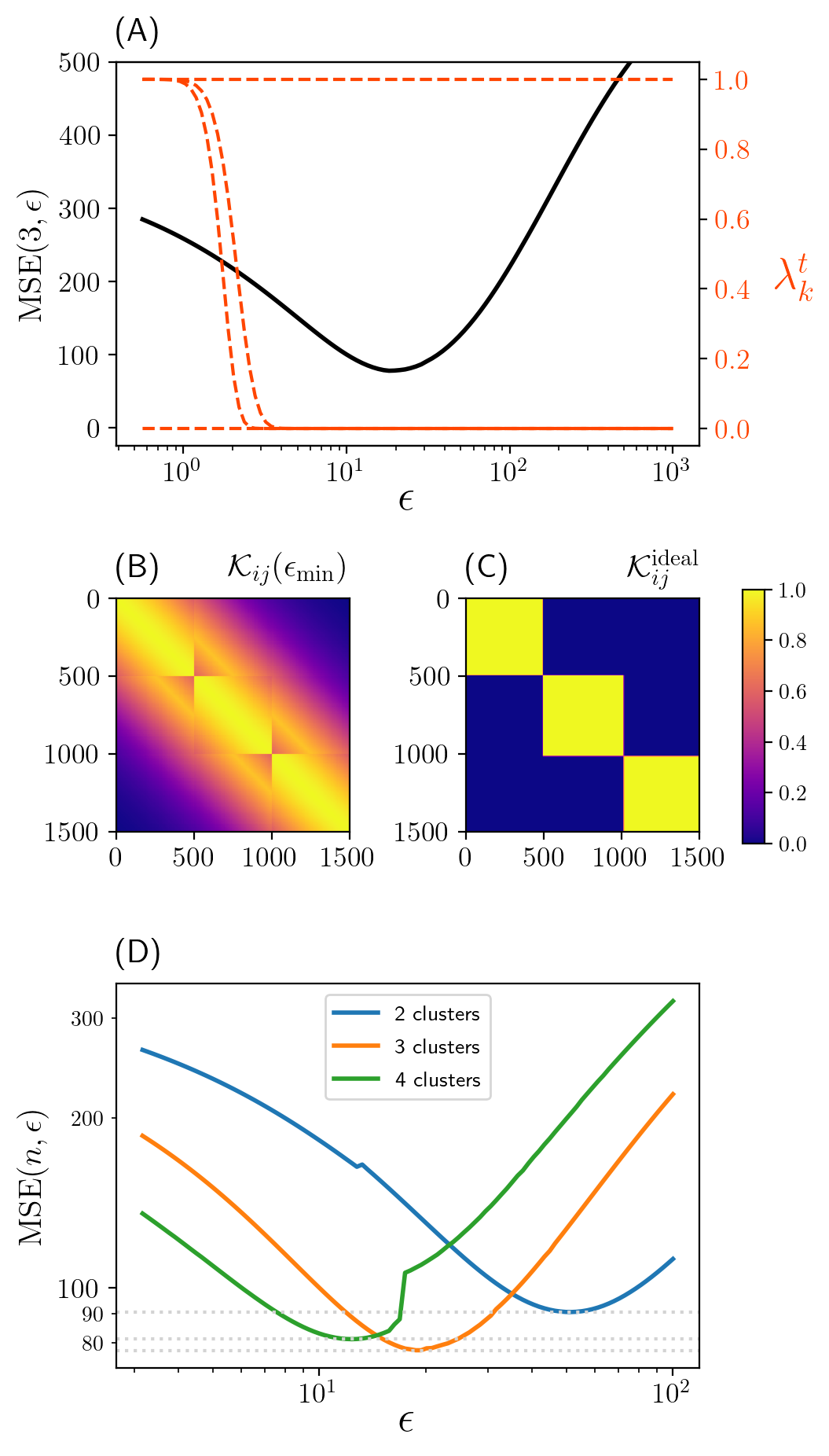}
	
	\caption{(A) The augmented MSE [black] from Equation \ref{eq:mse} of the spin data in Figure \ref{fig:diffmap}(A) as a function of resolution given 3 clusters.  Also plotted is the five largest transition probability eigenvalues [red] (Equation \ref{eq:eigendef}).  The minimum of the MSE can be found with standard computer routines, and allows a program to automatically select $ \epsilon $ without an experimenter choosing the hyperparameter.  (B) The similarity matrix when the MSE is minimized.  The resolution parameter is much larger here than in Figure \ref{fig:diffmap}(C), resulting in a more uniform matrix.  (C)  The `ideal' similarity which the heuristic tries to match. (D) The error for different numbers of k-means centroids on a log-log scale.  The error reaches its smallest possible value for $ n=3$ and $ \epsilon=18.2 $.  Based on the heuristic, those are the values with which diffusion maps is performed.  The lack of discontinuities in the $ n=3 $ curve may also be evidence for the proper number of cluster centers.  The locations of the centroids smoothly change for $ n=3 $, suggesting the data is inherently triply clustered in diffusion space.  \label{fig:mse}}
\end{figure}

Figure \ref{fig:mse}(A) shows the adjusted MSE of the XY spin data as a function of the resolution hyperparameter ($n=3$) and has a clearly defined optimum.  Minimizing the MSE as a function of $ \epsilon $ is a routine task for a computer.  Note this value for $ \epsilon $ in general exceeds the value that corresponds to the expected number of degenerate eigenvalues close to one.  The objective of the heuristic is to make the (normally sparse) similarity matrix very dense.  Figure \ref{fig:mse}(B) features a smeared similarity when the MSE is minimized.  The ideal similarity is displayed in Figure \ref{fig:mse}(C) for reference.  In addition to the resolution hyperparameter, the ideal number of k-means clusters can be automatically determined by the smallest optimized MSE as a function of $ n $.  Figure \ref{fig:mse}(D) features the MSE for $ n_{\text{clusters}} = \{2,3,4\} $.  The error is extremized for $ n=3 $; this is expected from the three distinct winding numbers present in the data.  Once this heuristic is established, a computer program can efficiently determine both the resolution and number of k-means clusters.  This process is aided by the fact that there are natural bounds for the scale of $ \epsilon $ related to the shortest and largest distance between each of the points in the original dataset.  The final result is a fully programmatic approach to perform diffusion maps on quantum states.  More details are given in Section \ref{sec:results}.

\section{Results}\label{sec:results}

\subsection{Haldane Model}

We first test our procedure on the Haldane model of graphene, a two dimensional array of carbon atoms in a honeycomb lattice with nearest and next nearest neighbor hopping \cite{gj6}. This model realizes the integer quantum Hall effect in graphene in the absence of a Landau level spectrum by adding a magnetic phase to the second nearest neighbor hopping parameters such that the net flux per plaquette is zero. The Haldane Hamiltonian in momentum space is \cite{gj6}
\begin{gather}
H(\mathbf{k}) = \mathbf{d}(\mathbf{k}) \cdot \boldsymbol{\sigma} \label{eq:bloch} \\ 
d_x(\mathbf{k}) = \cos(\mathbf{k} \cdot \mathbf{a}_1) + \cos(\mathbf{k} \cdot \mathbf{a}_2)+ 1 \\
d_y(\mathbf{k}) = \sin(\mathbf{k} \cdot \mathbf{a}_1) + \sin(\mathbf{k} \cdot \mathbf{a}_2) \\
\begin{split}
d_z(\mathbf{k}) = M + 2t_2 \sin(\phi) [ &\sin(\mathbf{k} \cdot \mathbf{a}_1) -\sin(\mathbf{k} \cdot \mathbf{a}_2)\\ &-\sin(\mathbf{k} \cdot (\mathbf{a}_1 - \mathbf{a}_2))  ]
\end{split}
\end{gather}
which is the Bloch vector form.  Here $ \mathbf{a}_1, \mathbf{a}_2 $ are the 2D lattice vectors, $ t_2 $ is the next-nearest neighbor hopping parameter, $ M $ is an onsite energy term that breaks inversion symmetry, and $ \phi $ is the magnetic phase associated with second nearest neighbor hopping.  In general there is a nearest neighbor hopping term ($ t_1 $), but here it is set to unity.  The Hamiltonian parameters of interest are $ M $ and $ \phi $.

\begin{figure}[t]
	\centering
	\includegraphics[width=3.3in]{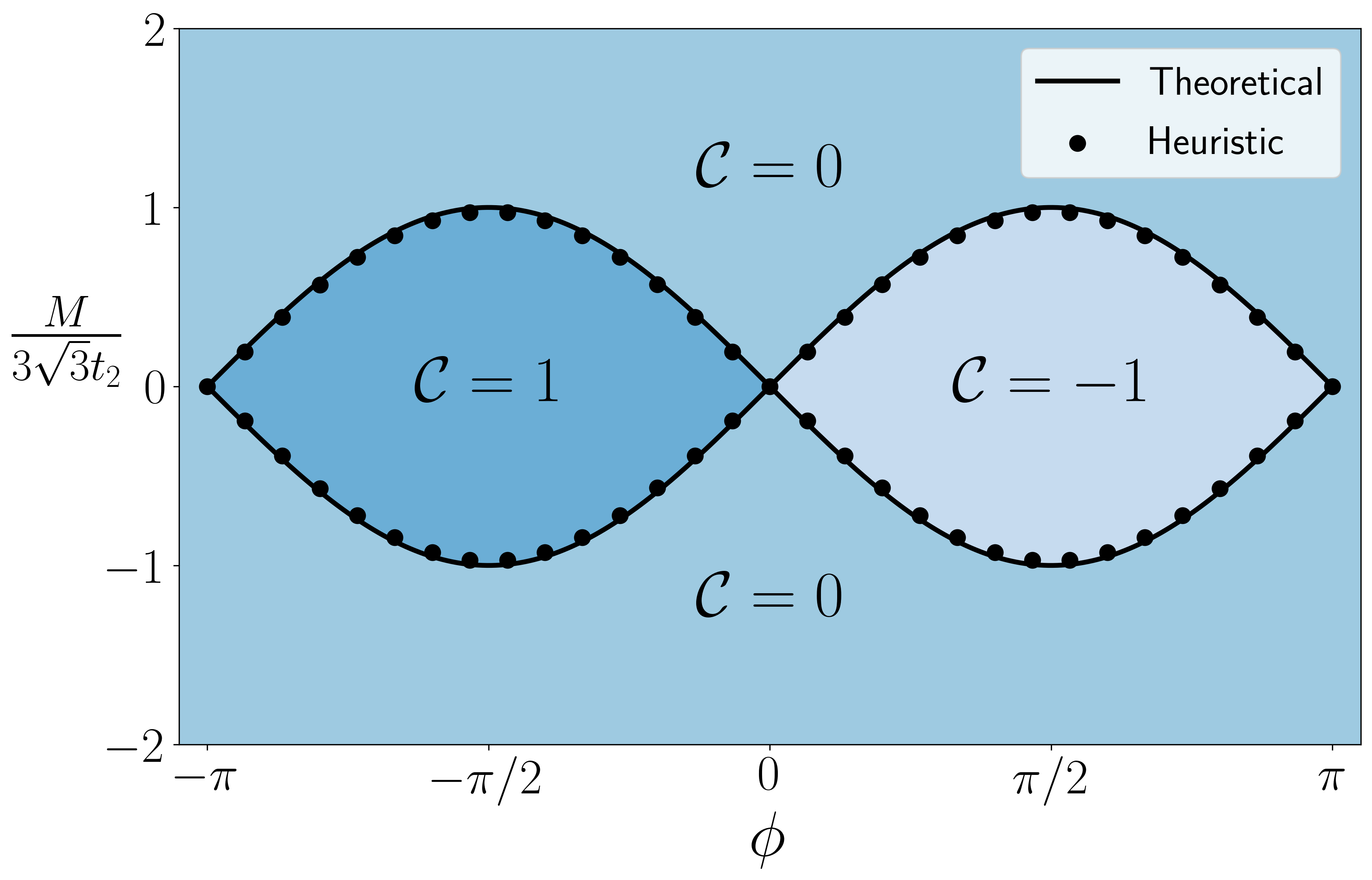}
	\caption{The learned phase boundaries (black dots) matches that of the known phase boundaries (black line) for the Haldane model.  The learned boundaries are in locations where the k-means labels of the Bloch vectors transformed by diffusion maps change.  Matter phases are identified by their Chern number and are color-coded here for extra visibility. \label{fig:haldane}}
\end{figure}

The topological invariant associated with the integer quantum Hall effect is the Chern number \cite{gj7}. For a Hamiltonian of the form in Equation \ref{eq:bloch} at half filling, the Hall conductivity is given by $ \sigma_{xy} = \frac{e^2}{h} \mathcal{C} $ where $ \mathcal{C} $ is the Chern number defined as

\begin{equation}
\mathcal{C} = \frac{1}{2\pi} \iint_{\text{BZ}} \text{d}k_x \text{d}k_y F_{xy}(\mathbf{k})
\end{equation}
with Berry curvature $F_{xy}(\mathbf{k}) = \frac{1}{2} \epsilon_{abc} \hat{\mathbf{d}}_a \partial_i \hat{\mathbf{d}}_b \partial_j \hat{\mathbf{d}}_c $ and $ \hat{\mathbf{d}}_a = d_a/|\mathbf{d}| $.  This integral is over the entire first Brillouin zone (BZ) which is spanned by the reciprocal lattice vectors.  The Chern number can be interpreted as the winding number of the map $ \mathbf{k} \rightarrow \hat{\mathbf{d}} $ from the BZ to a 2-sphere and is always an integer.  It is well understood \cite{gj6, carpentier, PhysRevB.95.144304, PhysRevB.95.184307, Caio2019} that as $ M $ and $ \phi $ are varied, the Haldane Hamiltonian undergoes phase transitions and can realize Chern numbers of 0, -1, and 1 (solid line in Figure \ref{fig:haldane}).

To test the heuristic's ability to draw the Haldane phase diagram, a dataset is composed of Bloch vectors from Equation \ref{eq:bloch}.  A single data point $ \boldsymbol{x}_i $ is a normalized Bloch vector $ \hat{\mathbf{d}} = \mathbf{d}/|\mathbf{d}| $ defined across the two-dimensional BZ mesh.  For a given value of $ \phi $ we build a dataset by uniformly sweeping $ M \in \left[-7, 7\right]$ in 1000 points.  The set $\{\boldsymbol{x}_i\}$ is then a collection of normalized Bloch vectors corresponding to a vertical slice of a phase diagram (Figure \ref{fig:haldane}).  The diffusion map resolution parameter and number of k-means centroids for a particular sample is determined via the procedure detailed in Section \ref{sec:eps-choice}.  K-means analysis automatically assigns cluster labels to each point of the transformed data $\{\boldsymbol{y}_i\}$.  Each of these clusters are understood to correspond to different topological phases.  Because the original dataset is composed of Bloch vectors that were sampled in an ordered manner (i.e. by monotonically increasing $ M $), the boundary between k-means clusters is interpreted as a Haldane phase transition.  On a phase diagram, the location in which the k-means labels change corresponds to the phase boundary.  This process is repeated for many choices of $ \phi $.  In Figure \ref{fig:haldane} we show a comparison between the boundary discovered by machine learning and that predicted by theory.  Without prior training, the algorithm can accurately determine the boundaries in the multi-phase system.

\subsection{Extended SSH Model}

Next we look at a generalized version of the Su-Schrieffer-Heeger (SSH) model describing electrons traveling across a 1D lattice with staggered lattice sites and next-next-nearest neighbor hopping (Figure \ref{fig:longrange-diagram}).  Like the Haldane model, the Hamiltonian in momentum space can be written in Bloch vector notation \cite{hsu}:

\begin{figure}[t]
	\centering
	\includegraphics[width=3.3in]{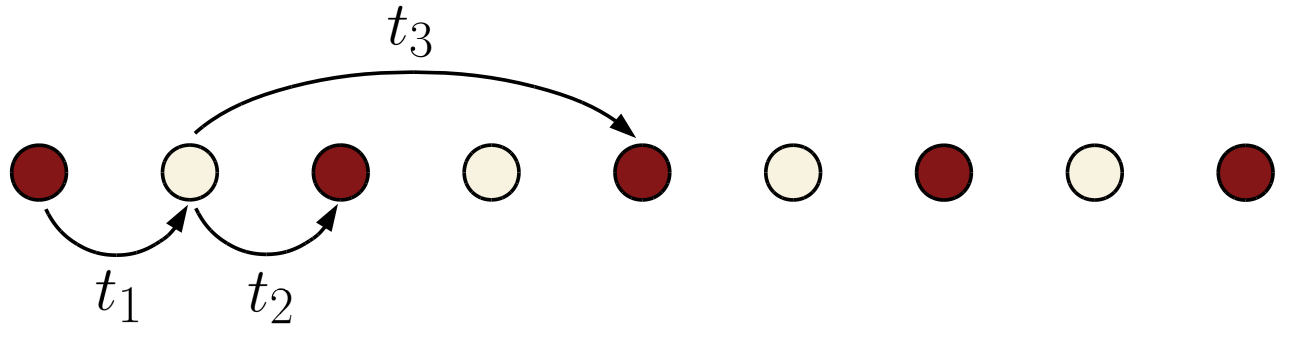}
	\caption{SSH Model with next-next-nearest neighbor hopping enabled, corresponding to the Hamiltonian of Equation \ref{eq:ssh}.\label{fig:longrange-diagram}}
\end{figure}

\begin{gather}
d_x(k) = t_1 + t_2 \cos(k) + t_3 \cos(2k) \nonumber \\
d_y(k) = t_2 \sin(k) + t_3 \sin(2k) \label{eq:ssh} \\
d_z(k) = 0 \nonumber
\end{gather}
When $ t_3 = 0 $ this reduces to the standard SSH model.  Phases of the SSH model differ by their winding number
\begin{equation}
\nu = \frac{1}{2\pi} \int_{-\pi}^{\pi} \text{d}k  \left( \hat{\mathbf{d}} \times \frac{\text{d}}{\text{d}k} \hat{\mathbf{d}}\right)_z
\end{equation}

\begin{figure}[t]
	\centering
	\includegraphics[width=3.3in]{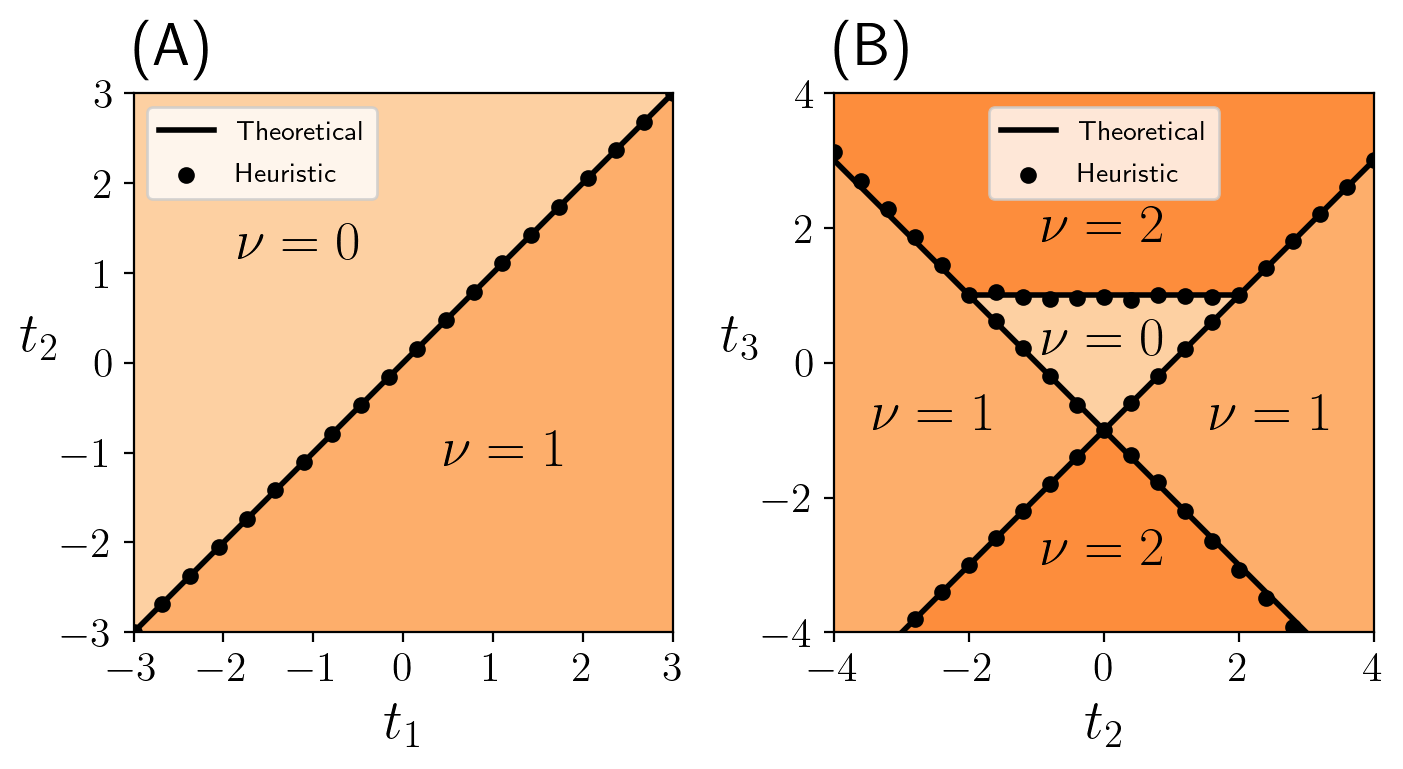}
	\caption{A) Learned phase diagram of the standard SSH model ($ t_3 = 0 $), matching theory to high precision.  (B) When next-next-nearest hopping is allowed ($ t_3 \neq 0, \; t_1 = 1 $), the heuristic is markedly less accurate but still close to the ground truth diagram.  Topological phases are identified even when they occupy a small volume of parameter space.  The ability of the heuristic to draw several boundaries simultaneously is also demonstrated.  \label{fig:ssh}}
\end{figure}

Again, the machine learning heuristic is tested against the known phase diagram (Figure \ref{fig:ssh}) by building a dataset through ordered changes in the model parameters.  Each data point $ \boldsymbol{x}_i $ is the normalized Bloch vector $ \mathbf{d} $ but now defined across 32 points in the one-dimensional BZ, $ k \in \left[ -\pi, \pi \right] $.  The automatic phase diagram process is applied by sweeping the relevant hopping parameters in Equation \ref{eq:ssh}.  In one experiment (Figure \ref{fig:ssh}(A)), datasets are composed of 1000 Bloch vectors for $ t_2 \in \left[-5, 5\right]$ on a grid of 20 $ t_1 $ values, while $ t_3 = 0$.  In another experiment (Figure \ref{fig:ssh}(B)), 1000 data points are generated where $t_3 \in \left[-5, 5\right]$ for slices of $ t_2 $.  Here $ t_1 = 1$ is fixed.  In replicating the phase diagram of the standard SSH model ($ t_3 = 0 $), the heuristic approach displays remarkable accuracy.  When long-range hopping is turned on ($ t_3 \neq 0$), the learning method still captures the true structure of the phase diagram despite the more complicated features in the data.  Even in regions where there are many phase boundaries $( |t_2| < 2 )$ the approach is successful.  This illustrates the heuristic's ability to identify topological order in small regions of parameter space.

\subsection{Triple Junction Quantum Ring Array - An Esoteric Case}

\begin{figure}[t]
	\centering
	\includegraphics[width=1.5in]{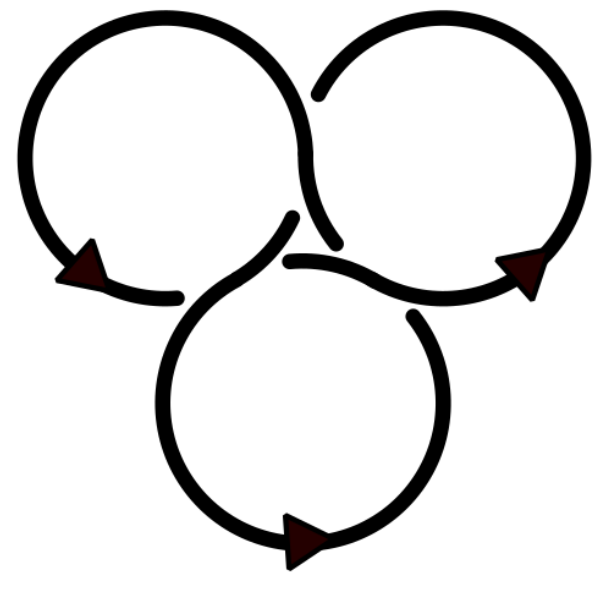}
	\caption{A schematic of the triple-ring junction, which can be understood as a single ring wrapped into the `trefoil' pattern.  At points where the wires cross there is a $ \delta $-function coupling that allows an electron to tunnel from one loop to the next. \label{fig:trefoil}}
\end{figure}

Lastly we investigate a system where a topological phase transition is known only through numerical analysis: single electrons on knotted systems of tunneling-coupled one-dimensional quantum rings.  One-dimensional quantum wires can be wrapped to form rings in several topologically distinct ways.  One such configuration is a single self-connected wire wrapped into a trefoil knot (Figure \ref{fig:trefoil}).  If one allows $\delta$-function tunnel coupling where the wire crosses itself, interesting effects of frustration and topology appear \cite{riggert}.  These effects are made richer by applying a magnetic field such that the tunnel-coupling matrix elements pick up a complex Aharonov-Bohm phase factor.

Experimentally, this phase factor can be changed by adjusting the strength of an applied magnetic field. As this phase is swept from $ -\pi $ to $ \pi $, it becomes energetically favorable for the wavefunction to accommodate this magnetic phase ``twist" by changing how the phase of the wavefunction itself winds about the knot in real space. This manifests in a spontaneous change in the winding number of the ground state wavefunction, defined as

\begin{equation}\label{eq:ringqnum}
n = -\frac{i}{2 \pi} \int z^* \frac{\partial z}{\partial s} ds 
\end{equation}
where
\begin{equation}
z = \frac{\psi}{|\psi|}
\end{equation} 
The exact coupling phase where this transition occurs is dependent on the magnitude of the tunnel coupling.

\begin{figure}[t]
	\centering
	\includegraphics[width=3.3in]{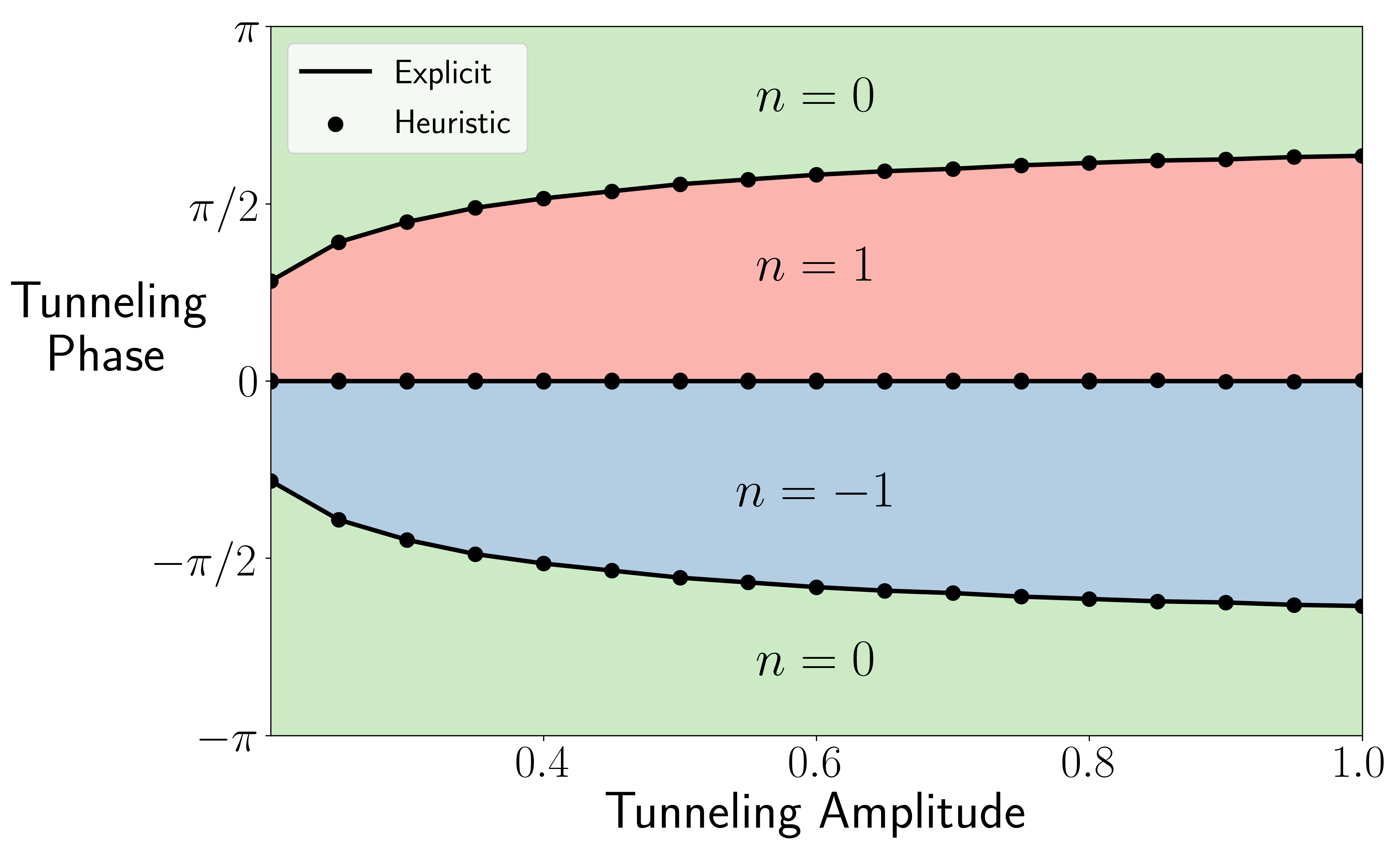}
	\caption{The phase diagram of the single-electron wavefunction across a triple-ring junction in Figure \ref{fig:trefoil}.  As one varies the real and imaginary parts of the $ \delta $-function coupling at the junctions, the wavefunction features stark shifts of its winding numbers as defined in Equation \ref{eq:ringqnum}.  The machine learning method (dotted line) is faithful to the explicit calculation of the winding number (solid line).\label{fig:rings}}
\end{figure}

In Figure \ref{fig:rings} we show the phase boundary determined by two numerical approaches.  The first is simply to explicitly calculate the phase winding in Equation \ref{eq:ringqnum} from the numerically calculated wavefunctions, for different values of coupling strength and magnetic field.  The second is to use the diffusion map heuristic where each data point $ \boldsymbol{x}_i $ is the ground state wavefunction normalized so that the phase at a particular position starts as pure real.  For different values of the tunneling amplitude, wavefunctions across the trefoil were calculated for many values of the tunneling phase.  The diffusion map approach identifies precisely the same phase boundaries even though it does not have the winding number explicitly coded into its algorithm.  This result reemphasizes the heuristic's ability to identify changes in generic topological structure.

\section{Conclusions}
Using the method of diffusion maps we have used computer learning to draw several very accurate phase diagrams of systems undergoing change in topological order.  These diagrams were determined with no human intervention beyond a supplied range of physicals parameters.  This process was used in models of different nature: two were well-established Hamiltonians defined in momentum space and the other was a calculated transition in a topological quantity in a lesser-known system.  The heuristic is successful even when many phase boundaries are encountered.  Furthermore, parts of the presented algorithm can be adjusted such as the distance metric to determine local similarity.  The result is a very general procedure to draw boundaries for TQPTs and beyond.  This approach could be used to investigate systems in which there is currently no known topological behavior.

The heuristic presented in this paper does have some limitations.  Firstly, there is no clear way to consider the absence of a phase transition.  This manifests because the MSE in Equation \ref{eq:mse} does not reach a finite value for $ n=1 $.  Because of this, the program will draw a phase boundary somewhere simply because it is forced to.  Secondly, it is possible the heuristic chooses an exceedingly large value for the resolution hyperparameter in cases where the similarity matrix is otherwise extremely sparse.  As a result, the true phase boundaries may not be accurately obtained.  This is a consequence of trying to match the similarity matrix with one that has sharply defined clusters in the objective function, Equation \ref{eq:mse}.  The resolution hyperparameter is tuned very high to get similarity values of 1 between many pairs of data points.  A simple diagnostic for this is to examine the sparseness of $ K_{ij} $ directly.  This is straightforward since even with multi-dimensional data the similarity matrix is two-dimensional.  If it is exceedingly sparse, alternative approaches may be necessary.

\section{Additional Note}
Code for the diffusion map and automatic resolution determination can be made available following the publication of this manuscript.

\bibliography{references}

\end{document}